\documentclass[prb,showpacs,showkeys,preprintnumbers,amsmath,amssymb,twocolumn]{revtex4-1}
\usepackage{makeidx} % Produzir indice remissivo
\usepackage{graphicx} % Incluir graficos no arquivo.
\usepackage{dcolumn} %Permitir alinhamento decimal em tabelas. Precisa do Array Package.
\usepackage{array} % Adicionar novas particularidades para as tabelas e vetores.
\usepackage{amssymb} %Permite o uso de sofisticados simbolos matematicos
\usepackage{amsmath}
\usepackage{textcomp}
\usepackage{multirow}
\usepackage{subfigure}
\usepackage{eucal}
\usepackage{mathrsfs}
\usepackage[utf8]{inputenc}
\usepackage[all]{xy}
\usepackage{float} %????
\usepackage{amsmath} %????
\usepackage{amsfonts} %????
\usepackage{bm} %????

\begin{document}

\title{Cs$_2$NaAl$_{1-x}$Cr$_x$F$_6$: A family of compounds presenting magnetocaloric effect}

\author{S. S. Pedro$^{1, 2}$ , J. C. G. Tedesco$^{3}$, F. Yokaichiya$^{4}$, P. Brand\~ao$^{5}$, A. M. Gomes$^{6}$, S. Landsgesell$^{7}$, M. J. M. Pires$^{8}$, L. P. Sosman$^{1}$, A. M. Mansanares$^{9}$, M. Reis$^{1}$ and H. N. Bordallo$^{3}$}
%\ead{sspedro@id.uff.br}

\affiliation{$^{1}$Instituto de F\'isica, Universidade do Estado do Rio de Janeiro, Rua S\~ao Francisco Xavier 524, 20559-900, Rio de Janeiro, RJ, Brazil}
\affiliation{$^{2}$Instituto de F\'isica, Universidade Federal Fluminense, Av. Gal. Milton Tavares de Souza, s/no, 24210-346, Niter\'oi, RJ, Brazil}
\affiliation{$^{3}$Niels Bohr Institute, University of Copenhagen, Universitetsparken 5, 2100, Copenhagen, Denmark}
\affiliation{$^{4}$Instituto de Pesquisas Energ\'eticas e Nucleares, Av. Lineu Prestes 2242, S\~ao Paulo, SP, Brazil}
\affiliation{$^{5}$CICECO and Chemistry Department, Universidade de Aveiro, Aveiro, Portugal}
\affiliation{$^{6}$Instituto de F\'isica, Universidade Federal do Rio de Janeiro, P.O. Box 68528, Rio de Janeiro, RJ, Brazil}
\affiliation{$^{7}$Helmholtz Zentrum Berlin, Hahn-Meitner Platz, 1, 14109 Berlin, Germany}
\affiliation{$^{8}$Instituto de Ci\^encia e Tecnologia - ICT, Universidade Federal dos Vales do Jequitinhonha e Mucuri, Diamantina, MG, Brazil}
\affiliation{$^{9}$Instituto de F\'isica Gleb Wataghin, Universidade Estadual de Campinas, Caixa Postal 6165, Campinas, S\~ao Paulo, Brazil}

\date{\today}

\begin{abstract}

In this paper we explore the magnetocaloric effect (MCE) of chromium-doped elpasolite Cs$_2$NaAl$_{1-x}$Cr$_x$F$_6$ (x = 0.01 and 0.62) single crystals. Our magnetization and heat capacity data show the magnetocaloric potentials to be comparable to those of garnets, perovskites and other fluorides, producing magnetic entropy changes of 0.5 J/kg$\cdot$K (x = 0.01) and 11 J/kg$\cdot$K (x = 0.62), and corresponding adiabatic temperature changes of 4 K and 8 K, respectively. These values are for a magnetic field change of 5 T at a temperature around 3 K. A clear Schottky anomaly below 10 K, which becomes more apparent when an external magnetic field is applied, was observed and related to the splitting of the Cr$^{3+}$ energy levels. These results hint at a new family of materials with potential wide use in cryorefrigeration.
\end{abstract}


\maketitle

Magnetic refrigerators and cryocoolers are promising devices based on the magnetocaloric effect (MCE), with applications including hydrogen liquefiers, high-speed computers and SQUID cooling. Thus, the quest for new materials which exhibit the MCE and promise  technological improvement has attracted much attention in recent years\cite{tishin, pecharsky, gomez, schapers}. Two important thermodynamic quantities characterize the MCE: the temperature change in an adiabatic process ($\Delta T_{ad}$) and the entropy change in an isothermal process ($\Delta S_T$) upon magnetic field variation. The latter is strictly related to the efficiency of a thermomagnetic cycle. The MCE is usually indirectly measured by using specific heat and magnetization data \cite{tishin}. 

Cryocoolers broke the 1\,K barrier for the first time using paramagnetic salts\cite{pecharsky}. However, despite their many applications over the intervening years, the low thermal conductivity of these salts is detrimental in adiabatic demagnetization applications\cite{pobell}, leading to the search for new materials with MCEs at lower temperatures. Among such materials is gadolinium gallium garnet and magnetic nanocomposites based on iron-substituted gadolinium gallium garnets\cite{mcmichael,levitin,shull,shull2}. Although some of these compounds have $\Delta S_T$ values comparable to the Gd-standard material, of about 3.4 J/kg$\cdot$K for a field variation of $\Delta \mu_0H$ = 1 T close to room temperature\cite{tang}, there are other systems that have a much larger entropy change, $\Delta S_T \approx$ 30 J/kg$\cdot$K under $\Delta \mu_0H$ = 5 T, at 5 K, as Gd$_3$Ga$_{5-x}$Fe$_x$O$_{12}$\cite{mcmichael}. Other successful examples consist of perovskite-type oxides\cite{zhong}. However, very few studies have 
so far reported on caloric effects in fluoride systems\cite{flerov2,fernandez,birk}, that being the aim of this paper.

From the point of view of optical and structural properties, Cs$_2$NaAl$_{1-x}$Cr$_x$F$_6$ elpasolite single crystals, which crystallize in the $R\bar{3}m$ space group, have been thoroughly investigated\cite{meyer,fargin,sosman,fonseca,bordallo,vrielinck2002,vrielinck2004,vrielinck2005,yoka,pedro}. The choice of the Cr$^{3+}$ ion as a doping impurity is justified by the fact that its $3d$ unfilled shell produces electronic transitions that increase the luminescent properties and the quantum yield of the material\cite{sosman,pedro,torchia2002,payne}. 

In this paper, we show the magnetocaloric potentials ($\Delta T_{ad}$ and $\Delta S_T$) in Cs$_2$NaAl$_{1-x}$Cr$_x$F$_6$ (x = 0.01 and 0.62) single crystals. We present single crystal X-ray diffraction studies to confirm the crystallographic parameters, along with EPR results to verify the local distortion caused by the Cr$^{3+}$ doping that occupies two nonequivalent octahedral sites. Magnetic susceptibility and specific heat measurements as a function of temperature and applied magnetic field  bring about an interesting Schottky-like anomaly at low temperatures under an applied magnetic field. From these data, we show that the $\Delta T_{ad}$ and $\Delta S_T$ of these materials are comparable to those of some well-known garnets and perovskite compounds, opening new doors for their application in cryorefrigeration.

Cs$_2$NaAl$_{1-x}$Cr$_x$F$_6$ single crystal samples of about $3\times 2 \times 1$\,mm$^{3}$ containing x = 0.01 and 0.62 of Cr$^{3+}$ doping were grown by hydrothermal techniques, using the temperature gradient method\cite{sosman}.

Single crystal X-ray diffraction data of the samples were collected on a Bruker Apex II CCD-based diffractometer using graphite monochromatized Mo-K$\alpha$ radiation. We also performed Neutron Activation Analysis experiments at the BER II in Berlin, Germany, on the x = 0.62 sample in order to check the compound's composition. EPR measurements on the x = 0.01 sample were performed in a commercial Varian E-12 spectrometer. Data were obtained in the X-band at room temperature.

Using a Quantum Design commercial PPMS, heat capacity measurements with fields of 0, 5 and 10 T were performed with the dominant face of the sample perpendicular to the magnetic field direction. Magnetization curves \textit{vs.} temperature at 0.1 T and \textit{vs.} applied magnetic field at several temperatures were acquired using commercial Quantum Design SQUID equipment. 

The crystal structure of Cs$_2$NaAl$_{1-x}$Cr$_x$F$_6$ was verified by single crystal X-ray diffraction. For the structure refinement, it was assumed that the Cr$^{3+}$ ions are statistically distributed in the Al sites \cite{fonseca}. Following this procedure, the goodness of fit values were 1.166 and 1.151 for x = 0.01 and 0.62 samples, respectively. Both samples were found to crystallize in the rhombohedral structure previously reported for similar compounds \cite{bordallo,bordallo2,babel}. The crystal structure contains two non-equivalent octahedral sites, where one of these sites is formed by AlF$_6$ octahedra sharing faces with two NaF$_6$ octahedra (here called S1 site), while the other site is composed of one AlF$_6$ unit sharing corners with six NaF$_6$ units (S2 site). 

The magnetic susceptibility data ($\chi = M/\mu_0H$, not shown) for the x  = 0.01 sample obtained at 0.1 T as a function of temperature (with applied magnetic field perpendicular to the dominant face of the sample) was measured. In order to eliminate the temperature independent diamagnetic contribution \cite{reis}, $|d\chi/dT|^{-1/2}$ was calculated as a function of temperature. As expected for magnetically isolated ions, this response showed a Curie-Weiss behavior, with a paramagnetic Curie temperature of $\theta_p$ = 0.30(9) K and an effective moment of $p_\textrm{eff}$ = 5.81(3) $\mu_B$/dopant-ion, which is much higher than the theoretical one for systems containing Cr$^{3+}$ in the S = 3/2 spin state ($p_{\textrm{eff}}$ (Cr$^{3+}$) = 3.87 $\mu_B$, considering only spin). In order to understand the high value of the effective magnetic moment, we also performed EPR measurements in this sample.

The characteristics of the line splitting of the grouped central EPR spectra (at several sample positions in the resonance cavity,  not shown) can be clearly assigned to Fe$^{3+}$ impurities. These grouped spectra display a signal in the magnetic field range of 0.25 to 0.45 T that was attributed to the presence of a Fe$^{3+}$ contamination in the sample, giving evidence that Fe$^{3+}$ and Cr$^{3+}$ occupy the same kind of site and such behavior can be seen in some previous works on similar compounds \cite{fargin,vrielinck2002,vrielinck2004,weil}. In this way, the apparent discrepancy with the magnetic moment obtained from Curie-Weiss fitting from inverse susceptibility data for the x = 0.01 sample is accounted by the small amount of Fe$^{3+}$ impurity detected by EPR, which shows a quite high effective magnetic moment of 5.92 $\mu_B$. 

The magnetic behavior of Cs$_2$NaAl$_{1-x}$Cr$_x$F$_6$ can be described by the following Hamiltonian \cite{weil}

\begin{equation}
\mathcal{H}=\mu_{B}(\vec{H}.\overleftrightarrow{g}.\hat{S})+D(\hat{S}^{2}_{z}-\frac{1}{3}\hat{S}^{2})
\end{equation}

which is characterized by two parameters: the $D$ parameter is related to the axial crystalline electrical field interaction and to the contribution to the orbital momentum of the excited states, while $\overleftrightarrow{g}$ is the spectroscopic splitting tensor \cite{abragam}. The resonance fields depend on the sample position relative to the applied field and they can be simulated choosing the best set of $\overleftrightarrow{g}$  and $D$ parameters.
 
As indicated by single crystal X-ray diffraction results, two different sites occupied by Cr$^{3+}$ ions were considered. We could not detect anisotropy in $\overleftrightarrow{g}$, $g = g_{\bot} = g_{//}$ = 1.95(1) for the S1 site and 1.97(1) for the S2 site. The $D$ parameter values for both sites were $D = -2540(50)$ cm$^{-1}$ (for S1 site) and $-3560(50)$ cm$^{-1}$ (for S2 site). By comparing our values to those reported in the literature, a reasonable agreement is found for the magnitude of this parameter \cite{fargin,vrielinck2002,vrielinck2004,vrielinck2005,brik}.

Turning to the magnetic susceptibility data obtained for the x = 0.62 sample, the calculated effective moment was $p_\textrm{eff}$ = 3.90(8) $\mu_B$/Cr$^{3+}$ ion, consistent for octahedral systems containing Cr$^{3+}$ in the S=3/2 spin state \cite{bizo,coste}. The negative paramagnetic Curie temperature extracted from fitting ($\theta_p = -0.20(4)$ K) indicates a tendency of antiferromagnetic arrangement between the Cr$^{3+}$ ions when the Cr$^{3+}$ concentration increases.

\begin{figure}[!htb] 
\begin{center}
\includegraphics[width=0.4\textwidth] {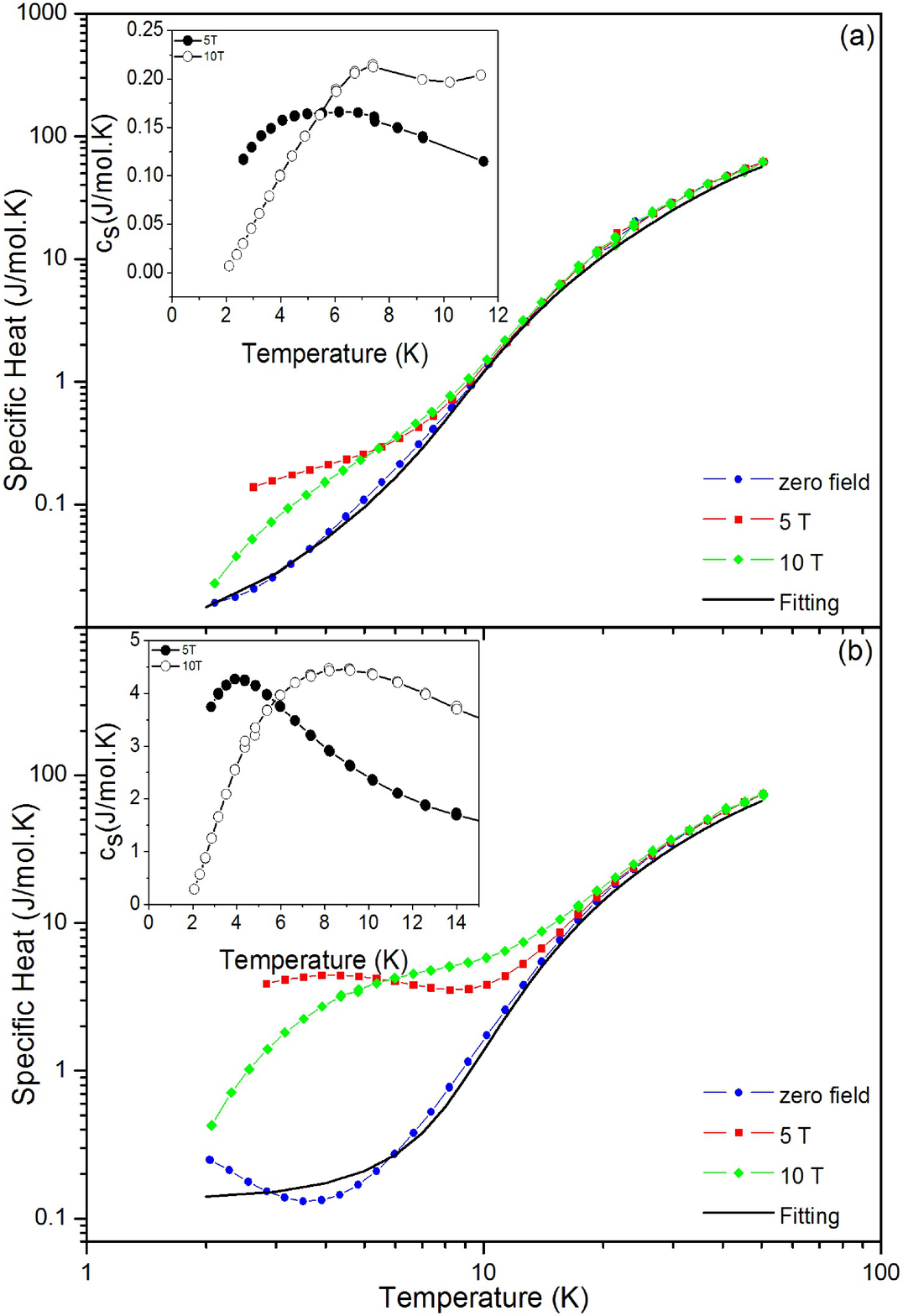}
\caption{(color online) Specific heat as a function of temperature and magnetic field of the system Cs$_{2}$NaAl$_{1-x}$Cr$_{x}$F$_{6}$ doped with (a) x = 0.01 and (b) x = 0.62 of Cr$^{3+}$. Blue data were obtained at zero magnetic field, red data at $\mu_0H$ = 5 T and green data at 10 T. The black line is the fit obtained from zero field data, considering only the lattice contribution to the specific heat. The insets are the estimated magnetic (Schottky) contribution to the specific heat for the sample at magnetic fields of 5 T (filled points) and 10 T (open points).}
\label{Figure1}
\end{center}
\end{figure}

Specific heat data as a function of temperature for applied fields equal to 0, 5 and 10 T are shown in Figure \ref{Figure1} for both samples, using a logarithmic scale for better visualization. One notices that above 10 K the magnetic field has no effect on the specific heat response, allowing us to conclude that the lattice contributions are dominant. However, if we turn to the analysis of the results with applied magnetic field for T$<$10 K, we observe a clear increase of the signal when the field is applied. We assign this change to the manifestation of the Schottky effect, which is an anomaly related to the splitting of the crystal field levels of the transition metal  inserted in the lattice \cite{evangelisti}. Such effects are often observed in rare-earth doped materials \cite{gofryk}, but also they can occur in systems with transition metal ions \cite{lazaro,affronte}. It is worth noticing that below 6 K the specific heat capacity increases with applied magnetic field in the interval from 0 to 5 T, 
and decreases for $\mu_0H >$ 5 T. 

In order to analyse and to separate the magnetic and the lattice contributions to the specific heat, we constructed a fitting routine based on specific heat data at zero field \cite{kholer,xie}. Two curves were generated: one considering the specific heat formulation according to the Debye theory \cite{kittel}, $c_D$; and the other considering the Einstein model \cite{kittel}, $c_E$. Although the Debye theory is more realistic than the Einstein model, the latter was also considered in the fitting procedure because it models more accurately the optical modes, while the Debye term makes better account of the acoustic modes \cite{kholer}. Using only the zero-field $c_p$ data, a fitting was performed with a weighting factor set as a free parameter. The black line in Figures \ref{Figure1}(a) and (b) represents the theoretical fit with equation $c_{latt} = 0.65c_D + 0.35c_E$ for the x = 0.01 sample, while for the x = 0.62 sample the best-fit equation was $c_{latt} = 0.49c_D + 0.51c_E$. It can be seen that the 
fitting to the experimental data obtained at zero field for the x = 0.01 data is better that for the x = 0.62 data.

From the fitting procedure, the Debye ($\theta_D$) and Einstein ($\theta_E$) temperatures were also obtained, with  $\theta_D$ = 230 K and $\theta_E$ = 73 K for the x = 0.01 sample, and $\theta_D$ = 241 K and $\theta_E$ = 80 K for the x = 0.62 sample. From low temperature luminescence experiments it is known that the vibrational lines are not shifted in relation to the zero-phonon line when the doping level is increased \cite{pedro}, indicating that the Cr$^{3+}$ impurity concentration does not have significant influence on the lattice vibrational modes \cite{kholer}. In this way, it can be seen that the difference between the values of $\theta_D$ and $\theta_E$ for both concentrations is not enough to cause a significant change in the lattice vibrational modes.

\begin{figure}[!htb]
\begin{center}
\includegraphics[width=0.4\textwidth] {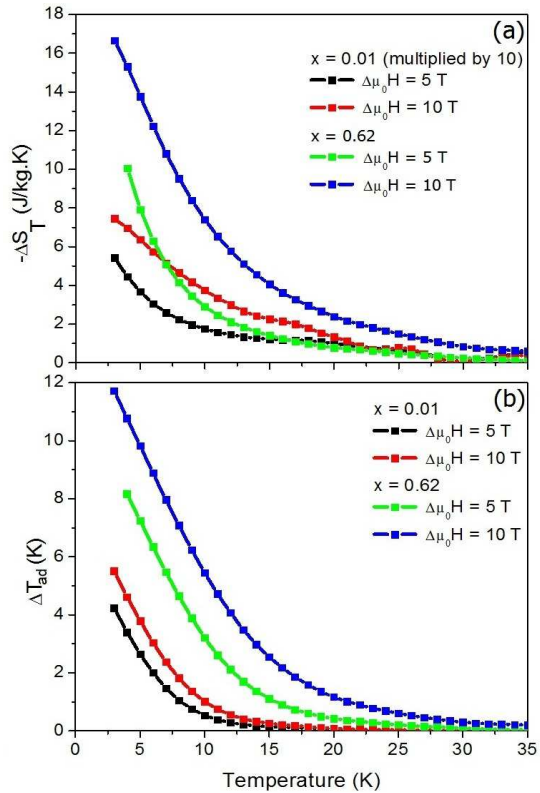}
\caption{(color online) Entropy changes (a) and temperature changes (b) of the system Cs$_{2}$NaAl$_{1-x}$Cr$_{x}$F$_{6}$ doped with x = 0.01 (black and red data) and x = 0.62 (green and blue data) due to variation of the magnetic field ($\Delta\mu_0H$ = 5 T and 10 T). For better visualization, $\Delta S_T$ for the 1\% sample was multiplied by 10.}
\label{Figure2}
\end{center}
\end{figure}

The Schottky contribution to the specific heat was obtained by subtracting the fitting curve from specific heat data obtained at non-zero fields\cite{kholer}. This contribution can be seen in the insets of Figure \ref{Figure1}. The shape of these curves indicates a fast rise on the lower temperature side, followed by a smooth fall on the higher temperature side. Additionally, the temperature where the maximum is observed increases with applied magnetic field and doping concentration. Similar behavior has been observed in the Ba$_{8-x}$Eu$_x$Ge$_{43}\Box_3$ system exhibiting the Schottky anomaly at low temperatures \cite{kholer}.

In some cases, the Schottky anomaly can be modeled by assuming a two levels system. However, for transition metal ions, the partially filled $d$ shell is the outer shell, and therefore very sensitive to crystal field effects. Thus, the Schottky anomaly observed in this Cr$^{3+}$-doped systems can be directly related to the thermal population of the doublets S = 3/2 and S = 1/2 where the ground electronic state $^{4}$A$_2$($^{4}$F) splits due to crystal field effects \cite{lazaro}.

Of more importance is to consider that together with this anomaly, a MCE is clearly observed in this fluoride. From the well-known Maxwell relations, it is possible to calculate the entropy change ($\Delta S_T$) due a variation of field ($\Delta\mu_0H$) using the following expression:

\begin{equation}
 \Delta S_{T}(T,\Delta H)=\int^{H2}_{H1}\left(\frac{\partial M(T,H)}{\partial T}\right)_{H}dH
\end{equation}

Moreover, as Cs$_2$NaAl$_{1-x}$Cr$_x$F$_6$ does not show structural phase transitions in the temperature range investigated, we can approximate this expression as a summation using the M \textit{vs.} H data \cite{tishin}. Another important quantity that features in the magnitude of the MCE is the temperature change ($\Delta T_{ad}$) due to $\Delta\mu_0H$, which can be obtained through the $c_p$ data as follows\cite{tishin}:

\begin{equation}
 S(T)_{H=0}=\int^{T}_{0}\frac{C(T)_{H=0}}{T}dT+S(0)_{H=0}
\end{equation}

and

\begin{equation}
 S(T)_{H\neq0}=\int^{T}_{0}\frac{C(T)_{H\neq0}}{T}dT+S(0)_{H\neq0}
\end{equation}

where $S(0)_{H=0}$ and $S(0)_{H\neq0}$ are the entropies at 0 K. 

Using the equations above, we obtain the curves shown in Figure \ref{Figure2}. It can be seen that the shape of the $\Delta S_T$ curves for both samples is very similar, with the higher-field curves always having higher values, and the $\Delta S_T$ values for x = 0.01 sample being around ten times smaller than for the x = 0.62 sample. Additionally, as the temperature decreases, $\Delta S_T$ increases. The values of $\Delta S_T$ at 3 K are summarized in Table 1. These results are comparable to the values obtained for other paramagnetic fluoride systems  \cite{fernandez,birk}. 

\begin{table}[!htc] 
\begin{center}
    \label{table1}
    \caption{Magnetic entropy and temperature changes for the Cs$_2$NaAl$_{1-x}$Cr$_x$F$_6$ and their comparison to other systems.}
    \begin{tabular}{|c|c|c|c|}
    \hline
    Sample                         &  $\Delta\mu_0H$  & $\Delta S_T$           & $\Delta T_{ad}$ \\
                                   &      (T)               &  (J/kg.K)                             & (K)  \\
    \hline    
    Cs$_2$NaAl$_{0.99}$Cr$_{0.01}$F$_6$  &  5            & 0.5                             & 4.2 \\
    (at 3 K - this work)                        &  10                 & 0.7                             & 5.5 \\ 
    \hline
    Cs$_2$NaAl$_{0.38}$Cr$_{0.62}$F$_6$  &  5            & 10.0                            & 8.2\\
    (at 3 K - this work)                        &  10                 & 16.6                            & 11.7\\
    \hline
    Gd                             &  5                  & 10                              & 12\\
    (at 300 K) \cite{tishin}       &                     &                                 & \\
    \hline
    Gd$_3$Ga$_5$O$_{12}$           &  5                  & 25                              & -\\
    (at 5 K) \cite{mcmichael}      &                     &                                 & \\
    \hline
    Cd$_{0.9}$Gd$_{0.1}$F$_{2.1}$  &  5                  & 7.2                             & -\\
    (at 5 K) \cite{fernandez}      &                     &                                 & \\
    \hline
    H$_{48}$C$_{44}$N$_{6}$O$_{12}$F$_{45}$Cr$_2$Gd$_3$               &  5                 & 22                              & -\\
    (at 1 K) \cite{birk}          &                     &                                 & \\
    \hline
    ErAl$_2$                       &  5                  & 37                              & 10\\
    (at 13 K) \cite{vonranke}     &                     &                                 & \\  
    \hline
    DyNi$_2$                       &  5                  & 21                             & 8\\  
    (at 21 K) \cite{vonranke}     &                     &                                 & \\
    \hline
    \end{tabular}
\end{center}
\end{table}

Common materials that display large MCEs below 80 K are intermetallic systems such as ErAl$_2$ and DyNi$_2$, which show temperature changes of $\Delta T_{ad} \approx$ 10 K (at T = 13 K) and $\Delta T_{ad} \approx$ 14 K (at T = 21 K), under $\Delta\mu_0H$ = 5 T, respectively \cite{vonranke}. Now, turning to the analysis of the $\Delta T_{ad}$ results in the Figure \ref{Figure2}(b), it is clear that the shapes of these curves are similar to the $\Delta S_T$ curves. From the $\Delta T_{ad}$ curves and the values given in Table 1, we can estimate that the impressive values of $\Delta T_{ad}$ obtained can make this material a promising cryocooler.

This paper shows that  Cs$_2$NaAl$_{1-x}$Cr$_x$F$_6$ single crystals (x = 0.01 and 0.62) exhibit large values of temperature variation under magnetic field changes. Such an observation hints that this family of compounds is a promising material to be used in  magnetic cryorefrigeration.  The magnetocaloric potentials were calculated from the specific heat and magnetization measurements. The values of $\Delta S_T$ are comparable to other applied materials and the values of $\Delta T_{ad}$ are relatively large, as can be seen from Table 1.

As mentioned previously, adiabatic demagnetization of paramagnetic salts was the first method of magnetic refrigeration to reach temperatures significantly below 1 K. Although the $^{3}$He-$^{4}$He dilution refrigerator has replaced this technology in some devices, it nevertheless has the convenience of being a continuous refrigeration method, and  paramagnetic refrigeration still has some advantages. For lower temperatures, where dilution refrigerators can become inefficient, magnetic refrigeration is still required. In this case, the starting conditions for magnetic refrigerators can be easily achieved with simple dilution refrigerators and superconducting magnets \cite{pobell}. Another physical disadvantage of paramagnetic salts is their low thermal conductivity. Because of this limitation, paramagnetic intermetallics and other compounds (like perovskites and garnets) have been studied and attracted some attention with respect to their magnetocaloric properties \cite{pecharsky}. Taking this into account 
together with the values of $\Delta S_T$ and $\Delta T_{ad}$ for our samples, we believe that the results found here can be an important contribution to  research on magnetic refrigeration.

\hspace{0.5cm}

We thank Pedro von Ranke (UERJ, Brazil) and Walter Kalceff (UTS, Australia) for fruitful discussions. J.C.G.T.'s participation in this work was funded by the Science without Borders Program. Access to  CICECO/Chemistry Department (Aveiro, Portugal), GPMR-UNICAMP (Campinas, Brazil), LMBT- UFF (Niter\'{o}i, Brazil), LBT-UFRJ (Rio de Janeiro, Brazil), BERII facilities and LaMMB MagLab (Berlin, Germany) are gratefully acknowledged by all authors. Financial support was provided by  Proppi/UFF, FAPERJ, FAPESP, CAPES, CNPq and FINEP.


\begin{thebibliography}{00}


\bibitem{tishin} A.M. Tishin, Y.I. Spichkin The Magnetocaloric Effect and its Applications, 1st edition, Institute of Physics, Bristol, Philadelphia, 2003.
\bibitem{pecharsky} V. K. Pecharsky and K. A. Gschneidner Jr. J. Magn. Mag. Mat. \textbf{200} 44 (1999).
\bibitem{gomez} J. R. G\'omez \textit{et al.}, Ren. Sust. Energ. Rev. \textbf{17} 74 (2013). 
\bibitem{schapers} M. Sch\"{a}pers \textit{et al.} Phys. Rev. B \textbf{88} 184410 (2013).
\bibitem{pobell} F. Pobell, Matter and methods at low temperatures, Springer, 2007.
\bibitem{mcmichael} R. D. McMichael, J. J. Ritter and R. D. Shull, J. Appl. Phys. \textbf{73} 6946 (1993).
\bibitem{levitin} R. Z. Levitin \textit{et al.}, J. Magn. Mag. Mat. \textbf{170} 223 (1997).
\bibitem{shull} R. D. Shull, R. D. McMichael and J. J. Ritter, Nano. Mat. \textbf{2} 205 (1993).
\bibitem{shull2} R. D. Shull, IEEE Trans. Mag. \textbf{29} 2614 (1993).
\bibitem{tang} T. Tang \textit{et al.}, J. Mag. Mag. Mat. \textbf{222} 110 (2000).
\bibitem{zhong} Z. Wei, A. Chak-Tong and D. You-Wei, Chin. Phys. B \textbf{22} 057501 (2013).
\bibitem{flerov2} I. N. Flerov \textit{et al.}, Cryst. Rep. \textbf{56} 9 (2011). 
\bibitem{fernandez} A. Fernand\'ez \textit{et al.}, Mat. Chem. Phys. \textbf{105} 62 (2007).
\bibitem{birk} T. Birk \text{et al.}, Inorg. Chem. \textbf{51(9)} 5435 (2012).
\bibitem{meyer} G. Meyer, Prog. Solid St. Chem. \textbf{14} 141 (1982).
\bibitem{fargin} E. Fargin, B. Lestienne and J. M. Dance, Sol. St. Commm. \textbf{75} 769 (1990).
\bibitem{sosman} L. P. Sosman \textit{et al.}, Sol. St. Comm. \textbf{114} 661 (2000).
\bibitem{fonseca} R. J. M. Fonseca \textit{et al.}, J. Fluores. \textbf{10} 375 (2000).
\bibitem{bordallo} H. N. Bordallo \textit{et al.}, J. Chem. Phys. \textbf{115} 4300 (2001).
\bibitem{vrielinck2002} H. Vrielinck \textit{et al.}, Rad. Eff. Fed. Sol. \textbf{157} 1155 (2002).
\bibitem{vrielinck2004} H. Vrielinck, F. Loncke, F. Callens, P. Matthys and N. M. Khaidukov, Phys. Rev. B \textbf{70} 144111 (2004). 
\bibitem{vrielinck2005} H. Vrielinck \textit{et al.}, Phys. St. Sol. (c) \textbf{2} 384 (2005). 
\bibitem{yoka} L. P. Sosman, F. Yokaichiya and H. N. Bordallo, J. Mag. Mat. Magn. \textbf{321} 2210 (2009).
\bibitem{pedro} S. S. Pedro \textit{et al.}, J. Lum. \textbf{134} 100 (2013).
\bibitem{torchia2002} G. A. Torchia \textit{et al.}, Opt. Mater. \textbf{20} 301 (2002).
\bibitem{payne} S. A. Payne \textit{et al.}, IEEE J. Quant. Elec. \textbf{28} 1188 (1992).
\bibitem{bordallo2} H. N. Bordallo \textit{et al.}, J. Phys.: Condens. Matter \textbf{14} 12383 (2002).
\bibitem{babel} D. Babel and R. Haegele, Mat. Res. Bull. \textbf{8} 1371 (1973).
\bibitem{reis} M. S. Reis, Fundamentals of Magnetism, New York: Elsevier, 2013. 
\bibitem{weil} J. A. Weil and J. R. Bolton, Electron paramagnetic resonance: elementary theory and practical applications, Hoboken: Jonh Wiley \& Sons, 2007.
\bibitem{abragam} A. Abragam and B. Bleaney, Electron Paramagnetic Ressonance of Transition Ions, Oxford: Oxford University Press, 2012.
\bibitem{brik} M. G. Brik and N. M. Avram, J. Opt. Adv. Mater. \textbf{8} 102 (2006).
\bibitem{bizo} L. Bizo \textit{et al.}, Adv. Func. Mat. \textbf{18} 777 (2008).
\bibitem{coste} S. Coste \textit{et al.}, Sol. St. Chem. \textbf{162} 195 (2001).
\bibitem{evangelisti} M. Evangelisti \textit{et al.}, J. Mat. Chem. \textbf{16} 2534 (2006).
\bibitem{gofryk} K. Gofryk, \textit{et al.}, Phys. Rev. B \textbf{83} 064513 (2011).
\bibitem{lazaro} F. J. L\'azaro \textit{et al.}, J. Physique \textbf{49} (C8) 825 (1988).
\bibitem{affronte} M. Affronte \textit{et al.}, J. Phys. D.: Appl. Phys. \textbf{40} 2999 (2007).
\bibitem{kholer} U. Kh\"{o}ler \textit{et al.}, Physica B \textbf{378-380} 263 (2006).
\bibitem{xie} L. Xie, T. S. Su and X. G. Li, Physica C \textbf{480} 14 (2012).
\bibitem{kittel} C. Kittel, Introduction to Solid State Physics, Jonh Wiley \& Sons, 2005.
\bibitem{vonranke} P. J. von Ranke, V. K. Pecharsky and K. A. Gschneidner, Phys. Rev. B \textbf{58} 12110 (1998).


\end{thebibliography}
\end{document}